\documentclass[12pt,amsmath,amssymb]{iopart}

\usepackage[T1]{fontenc} 
\usepackage{graphicx}
\usepackage{bm}
\usepackage{color}

\begin{document}

\title{Quasicondensation reexamined}

\author{Przemys\l{}aw Bienias$^1$, Krzysztof Paw\l{}owski$^1$$^2$, Mariusz Gajda$^3$$^4$, Kazimierz Rz\k{a}\.{z}ewski$^1$$^2$$^4$}

\address{$^1$ Center for Theoretical Physics, Polish Academy of Sciences, Aleja Lotnik\'ow 32/46, 02-668 Warsaw, Poland}
\address{$^2$ 5.Physikalisches Institut, Universit\"at Stuttgart, Pfaffenwaldring 57, 70550 Stuttgart, Germany}
\address{$^3$ Institute of Physics, Polish Academy of Sciences, Aleja Lotnik\'ow 32/46, 02-668 Warsaw, Poland}
\address{$^4$ Faculty of Mathematics and Sciences, Cardinal Stefan Wyszy\'nski University, ulica Dewajtis 5, 01-815, Warsaw, Poland}
\ead{pbienias@cft.edu.pl}
\date{\today}

\begin{abstract}
We study in detail the effect of quasicondensation.
We show that this effect is strictly related to dimensionality of the system.
It is present in one dimensional systems independently of interactions -- exists in repulsive, attractive or in non-interacting Bose gas in some range of temperatures below characteristic temperature of the quantum degeneracy.
Based on this observation we analyze the quasicondensation in terms of a ratio of the two largest eigenvalues of the single particle density matrix for the ideal gas.
We show that in the thermodynamic limit in higher dimensions the second largest eigenvalue  vanishes (as compared to the first one) with total number of particles as $\simeq N^{-\gamma}$ whereas goes to zero only logarithmically in one dimension.
We also study the effect of quasicondensation for various geometries of the system: from quasi-1D elongated one, through spherically symmetric 3D case to quasi-2D pancake-like geometry.

\end{abstract}

\pacs{67.85.Bc, 03.75.Hh, 05.30.Jp}
                                              
\maketitle

\section{\label{sec:intro}Introduction}
Properties of the interacting, harmonically trapped, ultracold gas are much more interesting in one dimension than in two and three dimensions.
In 2D and 3D a phase transition occurs whereas in 1D does not (surprisingly in quasi-1D systems two-step condensation is possible \cite{ketterle1997}).
Moreover, in symmetric 3D systems it was shown experimentally \cite{stenger1999,philliphs1999} that nearly all the way up to critical temperature the phase of the cloud is spatially uniform, equivalently: a coherence length of the system is equal to its size.
On the contrary, Petrov et al. have shown that the observation of the quasicondensate (condensate with fluctuating phase) is possible in 1D \cite{petrov2000}, 2D \cite{petrov2D2000} and very elongated 3D \cite{petrov2001} repulsive Bose gas.
Those predictions were confirmed experimentally \cite{dettmer2001, gerbier2003, hellweg2003}.
Recently density and phase properties of elongated systems were investigated in a number of experiments:  \cite{esteve2006,YangYang, armijo2011, manz2010}.
In spite of many theoretical attempts \cite{andersen2002, khawaja2003, decoherentToQuasi, cockburn2011} the theory of quasicondesates is still not as mature as the theory of condensates.
In this paper we shed a new light on the quasicondensation phenomenon.
In \cite{bienias2011} we stressed that shortening of the coherence length is not only the property of a repulsive gas but also of an attractive one. 
There exists a direct connection between the quasicondensation and the spectrum of a one-body density matrix
\footnote{In the description of partially coherent light spatial coherence modes are used also as eigenfunctions of the first order correlation function of the light field. 
See for instance  B. Saleh \textit{Photoelectron Statistics: With Applications to Spectroscopy and Optical Communication} (Springer-Verlag, New York, 1978)
 }.
More precisely, the quasicondensation occurs when occupation of more than one eigenmode is comparable with the total population
 \cite{penrose1956}.
In some sense it is similar to the "fragmented" spinor condensate \cite{leggett2001}.
In Section \ref{sec:inter} we show that properties of one dimensional Bose gas are similar regardless the sign of the interaction.
In Section \ref{sec:ideal} we explore coherence properties of the ideal gas in 1D, 2D, 3D and with arbitrary ratio of the trap frequencies.

\section{\label{sec:inter}Interacting gas}
Firstly, we study a one dimensional, weakly interacting Bose gas confined in a harmonic trap.
Thus, our Hamiltonian of the one dimensional Bose gas has a form:
\begin{eqnarray}
\nonumber H&=& \int \hat{\Psi}^{\dagger }(x)\left( \frac{p^2}{2m}+\frac{1}{2}m\,\omega_0^2 x^2  \right) \hat{\Psi}(x) dx +\\
& & +\,\frac{g}{2} \int \hat{\Psi}^{\dagger }(x) \hat{\Psi}^{\dagger }(x) \hat{\Psi}(x) \hat{\Psi}(x).
\label{eqn:hamiltonian}
\end{eqnarray}
The Hamiltonian is a sum of the single particle oscillator energy with mass $m$ and angular frequency $\omega$ and a conventional contact interaction with the coupling constant $g$.
Throughout this paper we use the oscillator units of position, energy and temperature, $\sqrt{\frac{\hbar}{m \omega_0}} $, $\hbar \omega_0$ and $\frac{\hbar \omega_0}{k_B}$ respectively.
Hence a dimensionless coupling $g$ is in units of $\sqrt{\frac{\hbar^3\omega_0}{m}}$.
Almost all results are calculated for 1000 atoms.

At the beginning we study a one dimensional, repulsive, weakly interacting Bose gas confined in a harmonic trap.
Our results are for the canonical statistical ensemble, thus the temperature is a control parameter.
In previous works we showed that in a wide range of temperatures such a system can be efficiently described using the so called classical field approximation \cite{bienias2011, bienias2011attr}.
To analyze the coherence properties of the system we use the first-order correlation function defined as: 
\begin{eqnarray}
g_1(-x, x) & = \frac{\langle \Psi^*(-x) \Psi(x) \rangle}{\langle |\Psi(x)|^2 \rangle} \label{eqn:g1} \\
\nonumber & = \frac{\langle \sqrt{n(-x)}e^{-i\phi(-x)}\sqrt{n(x)}e^{i\phi(x)} \rangle}{\langle n(x)\rangle}.
\end{eqnarray}

\begin{figure}
\includegraphics{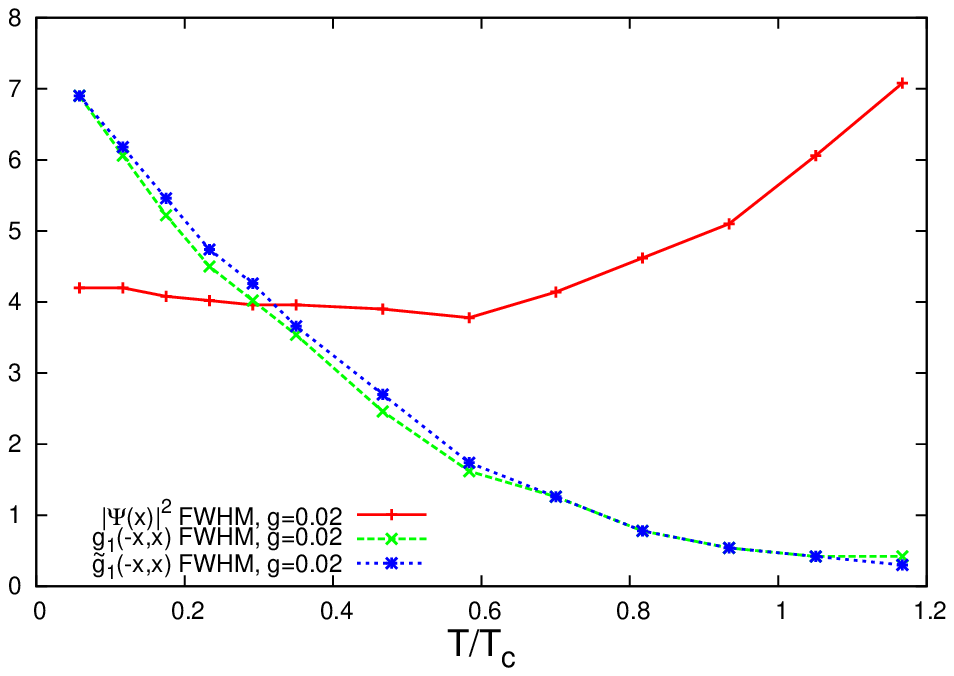}

\includegraphics{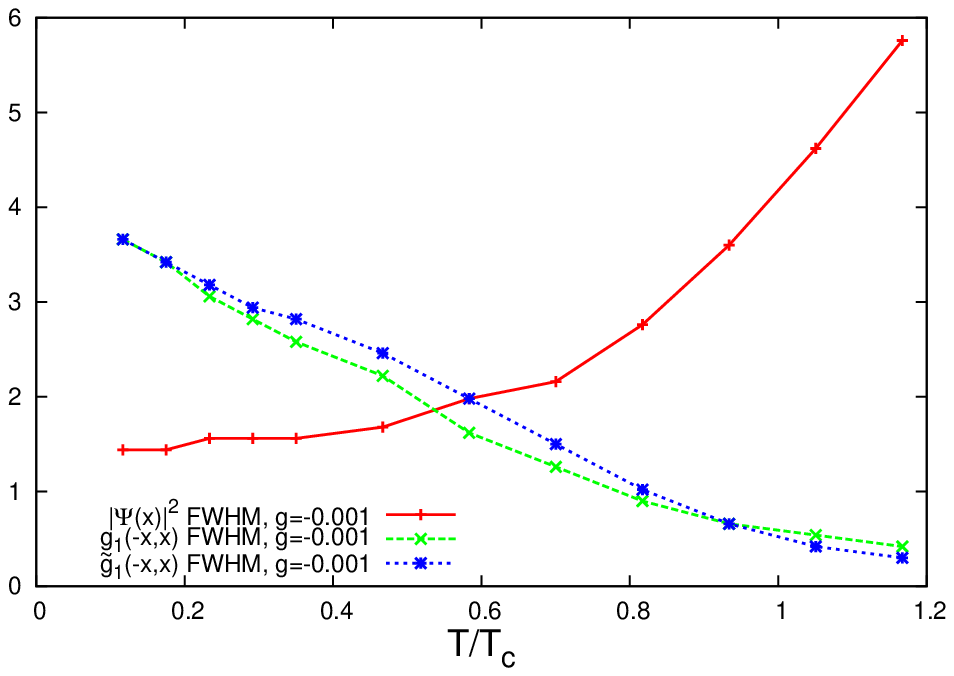}
\caption{(Color online) Correlation length, width of $\tilde{g}_1$ and size of the whole cloud.
All lengths are defined as full width at half maximum.
Results for (a) repulsive and (b) attractive gas.
$1000$ atoms considered.
}
\label{fig:g1_vs_T}
\end{figure}
Where $\langle \ldots\rangle$ is an ensemble average, $\phi(x)$ and $\sqrt{n(x)}$ are respectively: the phase and the absolute value of the wave function $\Psi(x)$ describing the whole system.
As a coherence length $l_{\phi}$ of the system we take the full-width at half maximum of the $g_1(-x,x)$.
Analogously the width of the atomic cloud is a full-width at half maximum of $|\Psi(x)|^2$.
In recent papers we showed that in both repulsive \cite{bienias2011} and attractive \cite{bienias2011attr} gas two regimes exist.
One between the zero temperature and $T_{ph}$ in which the size of the condensate is smaller than the coherence length.
The second one, above $T_{ph}$, in which the opposite condition occurs - a quasicondensate regime.
Such a situation is shown in Figure
\ref{fig:g1_vs_T} for weakly interacting repulsive gas.
To check how important are fluctuations of the phase in the behavior of $l_{\phi}$ we compared $g_1$ with: 
\begin{equation}
\tilde{g}_1 (-x, x) = \langle e^{-i\phi(-x)}e^{i\phi(x)} \rangle
\label{eqn:g1mod}
\end{equation}
Function $\tilde{g}_1$ unlike $g_1$ accounts for phase fluctuations only.
In Figure
\ref{fig:g1_vs_T} we see that in our regime of parameters $g_1(-x,x)$ is nearly the same as $\tilde{g}_1(-x,x)$, so density fluctuations do not contribute to the first-order correlation function regardless the sign of the interaction.
\begin{figure}
\includegraphics{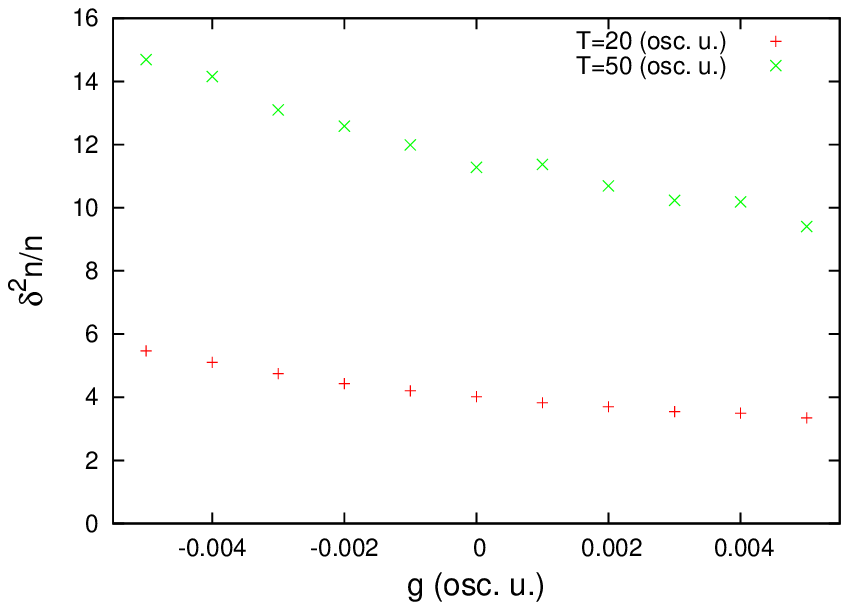}

\includegraphics{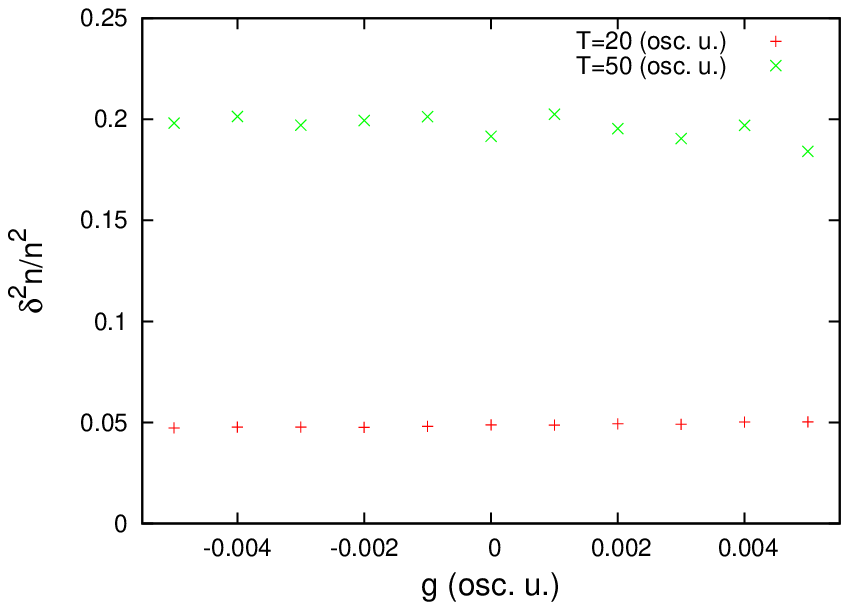}
\caption{(Color online) Behavior of the density fluctuations at the center of the trap versus interaction for 1000 atoms.
In (a) fluctuations divided by the density and in (b) divided by the square of density.}
\label{fig:densityFluct2}
\end{figure}
Moreover, density fluctuations are very similar for both attractive and repulsive cases.
In Figure
\ref{fig:densityFluct2} we show $\delta^2n/n$ (square of the density fluctuations divided by the density of atoms) at the center of the trap for changing $g$.
We see that for the repulsive gas the fluctuations are smaller than for the attractive one.
The reason is that for the repulsive case the mean-field interparticle interaction energy is positive and proportional to the density squared.
Because of this, the system prefers even distribution of the density rather than concentration of atoms.
In the attractive gas we have the opposite situation.
The interaction energy is lower for local bunching of atoms.
Nevertheless, we don't see any qualitative difference between positive and negative $g$, as the dependence is nearly linear.
What is more: the value of the fluctuations divided by the square of the density is the same for every $g$ what is presented in Figure
\ref{fig:densityFluct2}.
Finally, we see that coherence properties as well as density fluctuations are very similar for attractive and repulsive gas.
Therefore next, we take a closer look at the properties of the ideal gas.

\section{\label{sec:ideal}Ideal gas}
Detailed exact calculations are possible for the ideal gas. In the grand canonical ensemble results can be found in \cite{glauber1999,barnett2000,zyl2003}. 

We know that $\phi_i(x)$ in \eref{eqn:rho} are the eigenstates of harmonic oscillator.
In the canonical ensemble we used the recursion derived in \cite{wilkens} to calculate the partition function of $N$ atoms:
\begin{eqnarray}
\nonumber Z_0 & = 1,\\
Z_1(\beta) &=  \sum_{\nu}exp( - \beta \epsilon_{\nu}) \label{partFunc},\\
\nonumber Z_N(\beta)&= \frac{1}{N} \sum_{n=1}^N Z_1(n\beta) Z_{N-n}(\beta),
\end{eqnarray}
where $\epsilon_{\nu}$ is a single particle energy of the state $\nu$, $N$ is a number of atoms.
Knowing $Z_N$ we can calculate probability of finding $n$ atoms in a state of energy $\epsilon_{\nu}$.
Probability of finding at least $n$ atoms there is:
\begin{equation}
P_{\nu}^{\ge}(n|N)= e^{-n\beta \epsilon_{\nu}}\frac{Z_{N-n}}{Z_N}.
\end{equation}
Therefore:
\begin{equation}
P_{\nu}(n|N)= e^{-n\beta \epsilon_{\nu}}\frac{Z_{N-n}}{Z_N}-e^{-(n+1)\beta \epsilon_{\nu}}\frac{Z_{N-n-1}}{Z_N}.
\label{prob}
\end{equation}
Then we can find the average number of atoms in the $\nu$ mode:
\begin{equation}
N_{\nu}=\sum_n P_{\nu}(n|N)n.
\end{equation}
For deeper understanding of coherence properties lets go back to the  definition \eref{eqn:g1} of $g_1(-x,x)$.
We know that the one-body density matrix has the form:
\begin{equation}
\rho(x, y) = \langle \Psi^* (x)\Psi(y) \rangle = \sum_{i=0}\frac{N_i}{N}\phi_i^*(x)\phi_i(y),
\label{eqn:rho}
\end{equation}
where $N_i/N$ and $\phi_i(x)$ are eigenvalues and eigenvectors of $\rho(x,y)$.
Hence,
\begin{equation}
g_1(-x, x)= \frac{\sum_{i=0}\frac{N_i}{N}\phi_i^*(-x) \phi_i(x)}{\sum_{i=0}\frac{N_i}{N}\phi_i^*(x) \phi_i(x) }.
\label{eqn:g1_rho}
\end{equation}
We see a strong connection between the coherence length of the system and the spectrum of one-body density matrix.
Existence of a few comparable eigenvalues of the one-body density matrix signifies fragmented condensate \cite{gunn1998}.
It follows from the Penrose-Onsager criterion that the quasicondensate transition is caused by "sticking" of one-body density matrix eigenvalues 
- the situation when $N_0$ is comparable to $N_1$.

There are two important remarks which enable us to gain better intuition before further analysis.
First, when all atoms are in the condensate, this means that $N_0=N$, then $g_1(-x,x)$ is everywhere equal to one so we have a fully coherent system.
On the other hand if all 
eigenvalues are equal then $g_1(-x, x) \simeq \delta (x-x^{\prime})$ and coherence length is equal to zero.
\begin{figure}
\includegraphics{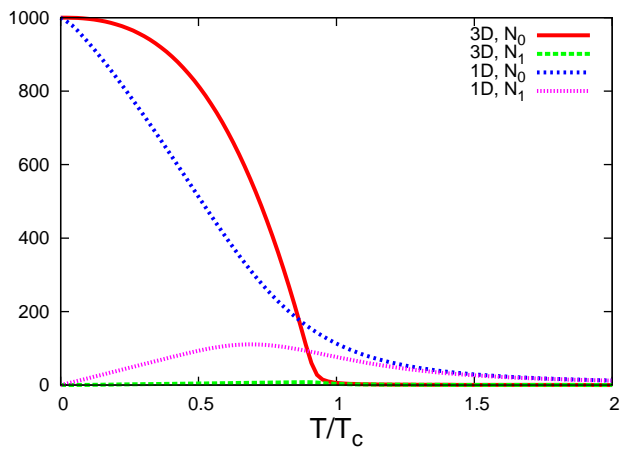}
\caption{(Color online) Comparison of 1D and 3D ideal Bose gas in a harmonic trap.
Populations of the ground state and the first excited state in both geometries for 1000 atoms are shown.
}
\label{fig:1Dvs3D}
\end{figure}
To show a notable difference in the form of density matrix spectrum in 1D and 3D for the ideal gas we present in Figure
\ref{fig:1Dvs3D}  a population of two lowest modes ($N_0$, $N_1$ respectively) as a function of temperature.
We see a qualitative difference between 1D and 3D.
In 3D $N_0$ is much bigger than $N_1$ all the way up to $T_c$.
Additionally the "sticking" of $N_0$ and $N_1$ takes place for almost vanishing $N_0$, $N_0/N \ll 1$, i.e.
almost at $T_c$.
On the other hand, in 1D we see the "sticking" for lower relative temperatures and, what is important, for much higher values of $N_0/N$.

Using the intuition gained, we analyze more precisely one, two and three dimensions.
The most intuitive way of comparing temperature effects for different regular geometries is to present temperature in $T_c$ units:
\begin{eqnarray}
\nonumber 1D: &  T_c& = \frac{\hbar \omega}{k_B}\frac{N}{\log(2 N)},\\
2D: &  T_c& = \frac{\hbar \omega}{k_B} \left(\frac{N}{\zeta(2)}\right)^{1/2} \label{Tc},\\
\nonumber 3D:  & T_c& = \frac{\hbar \omega}{k_B} \left(\frac{N}{\zeta(3)}\right)^{1/3}.
\end{eqnarray}
For 2D and 3D $T_c$ is equal to the critical temperature in thermodynamical limit.
In 1D the phase transition doesn't occur.
This is why only characteristic temperature is used \cite{ketterle1996}.
However to make a comparison for different aspect ratios with finite number of particles it is more convenient to keep a relative population of the condensate $N_0/N$ constant rather than the corresponding relative temperature.
\begin{figure}
\includegraphics{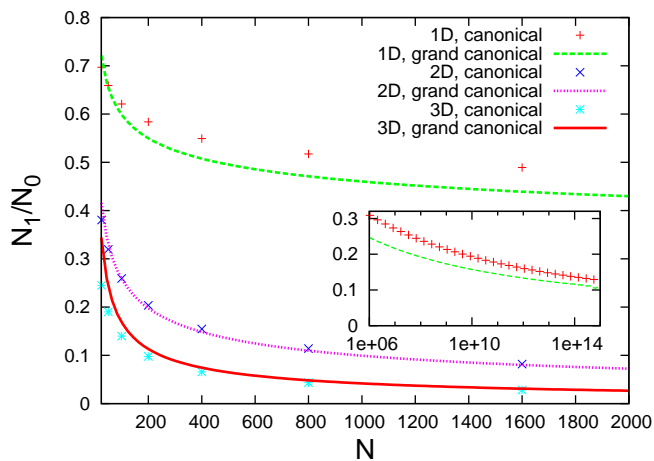}
\caption{(Color online) Ratio of populations of the ground state and the first excited state for 1D, 2D and 3D in canonical ensemble (points) for $N_0/N=0.2$.
Comparison with grand canonical results is presented (lines).
Note a good agreement in 2D and 3D.
Grand canonical results for 1D are calculated numerically using \eref{aeqn:averN}.
In the inset is shown, that the discrepancy between both ensembles in 1D exists even for huge number of atoms.
}
\label{fig:N1VsN}
\end{figure}

In Figure
\ref{fig:N1VsN} we show $N_1/N_0$ for different numbers of atoms in 1, 2 and 3 dimensions.
All results are for $N_0/N=0.2$.
In 1D `sticking' is definitely much stronger than in 2D and 3D.
Results in canonical ensemble are presented up to 1600 atoms - the number used in the experiment \cite{armijo2011}.
Because of numerical limitations the ratio of $N_1/N_0$ for large $N$ can be obtained in the grand canonical ensemble only.
We see a good agreement between canonical and grand canonical ensembles in crossover range of $N$ for 2D and 3D.

The agreement is worse in 1D.
This is because the ensembles are equivalent in the thermodynamic limit only.
Results of both ensembles (inset in Figure \ref{fig:N1VsN}) approach each other on the logarithmic scale.
Even for extremely large number of atoms the ratio $N_1/N_0$ in 1D is above $0.1$, thus one would expect quasicondensation even for very large one dimensional system.
Evidently, the finite size corrections in 1D systems are quite important.
The asymptotic behavior is not reached even for $N \simeq 10^{15}$ particles.

Moreover, the grand canonical approach allows also for finding the asymptotic value of the ratio of two dominant eigenvalues of the one-body density matrix.
The quantum partition function for a system of noninteracting bosons  in the grand canonical ensemble may be written as follows:
\begin{equation}
\mathcal{Z}(z, T) = \sum_{n_0=0}^{\infty}\sum_{n_1=0}^{\infty}\ldots \prod_{\lambda} z^{n_{\lambda}}\exp(-E_{\lambda} n_{\lambda} / k_B T),
\label{aeqn:part}
\end{equation}
where $E_{\lambda}$ is the energy of the single particle state, and $z = \exp\left({\mu/k_B T}\right)$  is a fugacity.
For atoms trapped in a harmonic potential with frequency $\omega$ we have 
in 1D: $E_\lambda = \hbar \omega \lambda$, 2D:  $E_\lambda = \hbar \omega (\lambda_x + \lambda_y)$ and in 3D: $E_\lambda = \hbar \omega (\lambda_x + \lambda_y + \lambda_z)$.
Knowing that $\langle N \rangle~=~z\frac{\partial} {\partial z} \ln \mathcal{Z}(z, T)$ we get an expression for the average number of atoms in the system:
\begin{equation}
N=\frac{z}{1-z} + \sum_{\lambda\neq 0}\frac{z e^{-\beta E_{\lambda}}}{1-z e^{-\beta E_{\lambda}}},  
\label{aeqn:averN}
\end{equation}
where we have extracted the term corresponding to the lowest energy $E_\lambda=0$.
In 2D and 3D we can safely set $z=1$ in the sum over all excited states \eref{aeqn:averN}.
Next, we can expand the formula \eref{aeqn:averN} considering only leading terms in a small parameter $\beta \hbar \omega$ \cite{castain}, 
it leads to the expression for the atom number:
\begin{equation}
N=\frac{z}{1-z}+\zeta(D)\left(\frac{k_B T}{\hbar \omega}\right)^{D},
\label{aeqn:3D}
\end{equation}
where $D=2~(3)$ in 2D~(3D).
Remembering that we are interested in results for the constant value $N_0=C N=\frac{z}{1-z}$ 
we get $z=\frac{C N}{1+C N}$ and $\frac{k_B T}{\hbar \omega} = \left( \frac{N(1-C)}{\zeta(D)} \right)^{1/D}$.
In the 1D in the thermodynamic limit we have
\begin{equation}
N=\frac{z}{1-z}-\ln(1-z) \frac{k_B T}{\hbar \omega}.
\label{aeqn:1D}
\end{equation}
Finally, we get that $\frac{k_B T}{\hbar \omega} $ is equal to $ \frac{N(1-C)}{\ln(C N+1)}$.
The asymptotic behavior of $N_1/N_0$ can be obtained from the relation: 
\begin{equation}
\lim_{N\to\infty} \frac{N_1}{N_0} = \frac{z e^{-\hbar \omega \beta}}{1- z e^{-\hbar \omega \beta}} \frac{1-z}{z}.
\end{equation}
Finally, in 1D we get the following scaling of the `sticking' ratio:
\begin{equation}
\lim_{N\to\infty} \frac{N_1}{N_0} \sim \frac{1}{\ln(N)},
\end{equation}
while in higher dimensions we have:
\begin{equation}
\lim_{N\to\infty} \frac{N_1}{N_0} \sim \frac{1}{N^{(D-1)/D}};
\label{aeqn:lim2D}
\end{equation}
that is
\begin{displaymath}
 \frac{1}{N^{1/2}} \quad\mathrm{for}\quad  D=2
\end{displaymath}
and
\begin{displaymath}
 \frac{1}{N^{2/3}} \quad\mathrm{for}\quad  D=3\,.
\end{displaymath}
The sticking ratio has totally different behavior in the asymptotic limit depending on the dimensionality of the system.
In higher dimensions it goes to zero with dimension-depend power while in 1D the decay of the sticking ratio is logarithmically slow.
This is the reason while in 1D system there is always a finite range of temperatures below the quantum degeneracy temperature where occupation of the higher orbitals of the one-body density matrix is relatively high.
\begin{figure}
\includegraphics{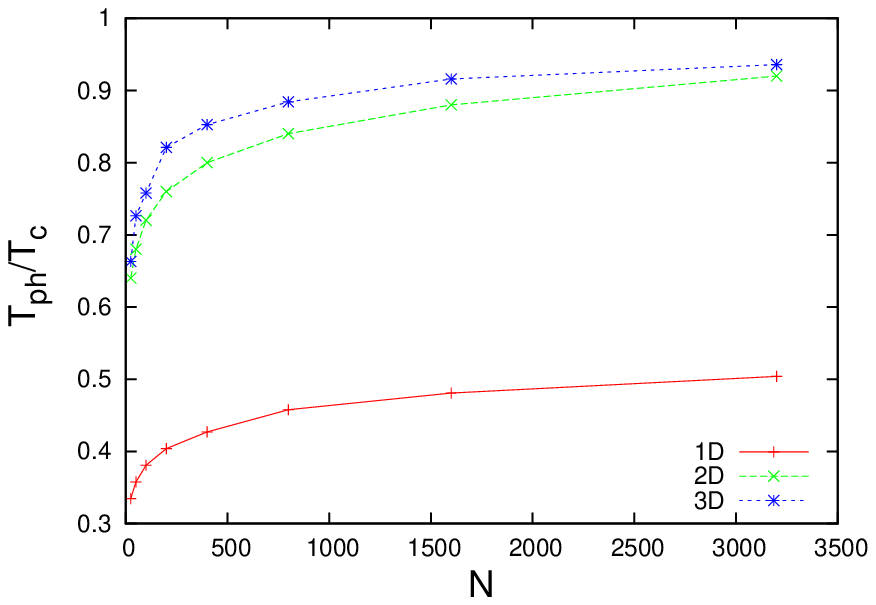}

\includegraphics{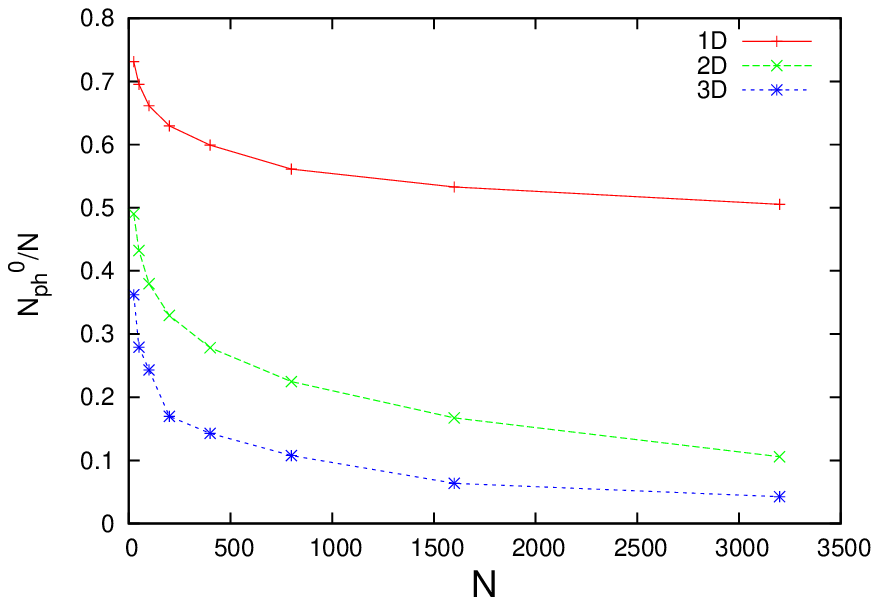}
\caption{(Color online) Parameters of the system at the point of the transition from true condensate to quasicondensate versus number of atoms.
In (a) temperature of the transition is presented, in (b) fraction of atoms in the ground state at the transition temperature.}
\label{fig:N0ph}
\end{figure}
The spectrum of one-body density matrix enables us to look at the dependence of $T_{ph}$ (the temperature when the width of the system is equal to the coherence length) as a function of $N$.
In Figure
\ref{fig:N0ph}(a) for better comparison we divided $T_{ph}$ by $T_c$ \eref{Tc}.
Once more a significant difference between 1D and higher dimensions is seen.
For 2D and 3D the temperature of quasicondensation is really near $T_c$ whereas for 1D the temperature $T_{ph}$ is much less than $T_c$.
Because of the ambiguity of the $T_c$ definition in 1D we looked at $N_{ph}^0$ (a number of atoms in the ground state at the temperature $T_{ph}$).
In Figure
\ref{fig:N0ph} (b) we show $N_{ph}^0/N$ as a function of $N$.
The ratio $N_{ph}^0/N$ is a fast decreasing function tending to zero in 2D and 3D, while for 1D it decreases much slower exceeding the value $0.5$ in the whole presented region.
\begin{figure}
\includegraphics{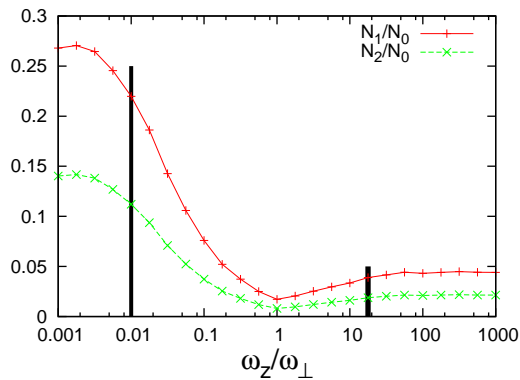}
\caption{(Color online) Relative number of atoms in the ground state and in the first excited state versus the aspect ratio $\omega_z/\omega_{\perp}$ for $N=1000$ and $N_0/N=0.4$.
The left vertical line denotes point when $k_B T_{ph} = \hbar \omega_{\perp}$ and the right one when $k_B T_{ph} = \hbar \omega_z$.
}
\label{fig:cross}
\end{figure}
Having analyzed 1D, 2D and 3D symmetric cases we study the ratio of $N_1/N_0$ and $N_2/N_0$ versus trap aspect ratio $\omega_Z/\omega_{\perp}$.
In Figure \ref{fig:cross} we present results for fixed: $N=1000$ and $N_0/N=0.4$.
This way we can analyze transition between different geometries of the system.
When $\omega_Z\ll\omega_{\perp}$ the system is nearly 1D, when $\omega_Z\approx\omega_{\perp}$ is nearly 3D, and when $\omega_z\gg\omega_{\perp}$ is nearly 2D.
Once more we see a clear difference between dimensions.
However, we do not observe any sharp transition between them.
The left vertical line corresponds to $k_B T_{ph} = \hbar \omega_{\perp}$ and the right to $k_B T_{ph} = \hbar \omega_z$, so to the situations when system can be considered respectively as one and two dimensional.
We see that these points agree with the beginning of the flattening of $N_1/N_0$ and $N_2/N_0$.
Moreover 2D case is much more similar to 3D than to 1D.
In experiments we never have exactly symmetric systems, however relatively broad range of aspect ratios around $\omega_z=\omega_{\perp}$ corresponds to systems which can be considered as 3D because of the flattening of $N_1/N_0$ in this region.
We checked that this region gets broader with increasing $N$.

Summarizing, we have shown that quasicondensation phenomenon is strongly related to the dimensionality of the system rather then to interactions.
Using the sticking ratio $N_1/N_0$ as the quasicondensation criterion we found the asymptotic behavior of this quantity in the ideal gas case.
The 1D asymptotic of the sticking ratio differs drastically from asymptotic in higher dimensions.
Vanishing of the second largest eigenvalue of the one-body density matrix is logarithmically slow in 1D.
Some decreasing of the correlation length in 2D and 3D can be also observed but this effect results from the finite number of atoms in the system and occurs just below the critical temperature.

\begin{ack}
\label{sec:acknowl}
This work was supported by Polish Government Funds for the years 2010-2012.
Two of us (K.P.and K.Rz.) acknowledge financial support of the project  "Decoherence in long range interacting
quantum systems and devices" sponsored by the Baden-W\"urtenberg Stiftung".
\end{ack}

\section*{References}
\bibliography{refsy}
\bibliographystyle{abbrv}

\end{document}